\documentclass[twocolumn,showpacs,prl,superscriptaddress,preprintnumbers]{revtex4}
\usepackage{graphicx}

\def\Re{{\rm Re}}
\def\beq{\begin{equation}}
\def\eeq{\end{equation}}
\def\bea{\begin{eqnarray}}
\def\eea{\end{eqnarray}}
\def\pepa{p_1\cdot p_2}
\def\pepb{p_3\cdot p_2}
\def\ubar{\bar{u}}
\def\slash#1{{\rlap{\hspace{.08em}/}#1}}

\begin{document}

\title{Two-photon exchange and elastic electron-proton scattering}

\author{P.\ G.\ Blunden}
\affiliation{Department of Physics and Astronomy, University of Manitoba,
Winnipeg, MB, Canada R3T 2N2}
\affiliation{Jefferson Lab, 12000 Jefferson Ave.,
Newport News, VA 23606}
\author{W.\ Melnitchouk}
\affiliation{Jefferson Lab, 12000 Jefferson Ave.,
Newport News, VA 23606}
\author{J.\ A.\ Tjon}
\affiliation{Jefferson Lab, 12000 Jefferson Ave.,
Newport News, VA 23606}
\affiliation{Department of Physics, University of Maryland, College Park,
MD 20742-4111}

\begin{abstract}
Two-photon exchange contributions to elastic electron-proton scattering
cross sections are evaluated in a simple hadronic model including the
finite size of the proton. The corrections are found to be small in
magnitude, but with a strong angular dependence at fixed $Q^2$. This is
significant for the Rosenbluth technique for determining the ratio of
the electric and magnetic form factors of the proton at high $Q^2$, and
partly reconciles the apparent discrepancy with the results of the
polarization transfer technique.
\end{abstract}

\pacs{25.30.Bf, 12.20.Ds, 13.40.Gp, 24.85.+p}

\maketitle

The electromagnetic structure of the proton is reflected in the Sachs
electric ($G_E(Q^2)$) and magnetic ($G_M(Q^2)$) form factors. The ratio
$R=\mu_p G_E/G_M$, where $\mu_p$ is the proton magnetic moment, has
been determined using two experimental techniques. The Rosenbluth, or
longitudinal-transverse (LT), separation extracts $R^2$ from the
angular-dependence of the elastic electron-proton scattering cross
section at fixed momentum transfer $Q^2$. The results are consistent
with $R\approx 1$ for $Q^2<6$~GeV$^2$~\cite{Arr03,Wal94}. However,
recent polarization transfer experiments at Jefferson Lab~\cite{Jon00}
measure $R$ from the ratio of the transverse to longitudinal
polarizations of the recoiling proton, yielding the markedly different
result $R\approx 1-0.135 (Q^2-0.24)$ over the same range in
$Q^2$~\cite{Arr03}, which exhibits nonscaling behavior. In this letter
we examine whether this discrepancy can be explained by a reanalysis of
the radiative corrections, in particular as they affect the LT
separation analysis.

Consider the elastic $ep$ scattering process $e(p_1) +
p(p_2)\rightarrow e(p_3)+p(p_4)$. The Born amplitude for one photon
exchange is given by
\beq
{\cal M}_0 = -i {e^2\over q^2} \ubar(p_3) \gamma_\mu u(p_1)\ 
\ubar(p_4) \Gamma^\mu (q) u(p_2),\label{m0}
\eeq
where the proton current operator is defined as
\beq
\Gamma^\mu(q) = F_1(q^2) \gamma^\mu + i {F_2(q^2)\over 2
M}\sigma^{\mu\nu}q_\nu,\label{current}
\eeq
$q=p_4-p_2=p_1-p_3$ is the four-momentum transferred to the proton
($Q^2\equiv -q^2>0$), $M$ is the proton mass, and $F_1$ and $F_2$ are
linear combinations of the Sachs form factors $G_E$ and $G_M$ (see
Eqs.\ (\ref{f1tosachs})--(\ref{f2tosachs})).

The resulting cross section depends on two kinematic variables,
conventionally taken to be $Q^2$ (or $\tau\equiv Q^2/4M^2$) and either
the scattering angle $\theta$ or the virtual photon polarization
$\epsilon = \left(1+ 2 (1+\tau) \tan^2{(\theta/2)}\right)^{-1}$. It can
be put in the form
\beq
d\sigma_0 = A \left(\tau G_M^2(Q^2) + \epsilon G_E^2(Q^2)
\right),\label{sigma0}
\eeq
where $A$ depends on kinematic variables. This expression is modified
by radiative corrections, expressed in the form $d\sigma = d\sigma_0
(1+\delta)$. Usually $\delta$ is estimated by taking the one-loop
virtual corrections of order $\alpha$ as well as the inelastic
bremsstrahlung cross section for real photon emission.

The LT separation technique extracts the ratio $(G_E/G_M)^2$ from the
$\epsilon$-dependence of the cross section at fixed $Q^2$. With
increasing $Q^2$ the cross section is dominated by $G_M$, while the
relative contribution of the $G_E$ term is diminished. Hence
understanding the $\epsilon$-dependence in the radiative correction
$\delta$ becomes increasingly important at high $Q^2$. By contrast, the
polarization transfer technique involves a ratio of cross sections, and
is not expected to show the same sensitivity to the
$\epsilon$-dependence of $\delta$~\cite{Afa01}.

The amplitude ${\cal M}_1$ for the one-loop virtual corrections can be
written as the sum of a ``factorizable'' term, proportional to the
Born amplitude ${\cal M}_0$, plus a remainder:
\beq
{\cal M}_1 = f(Q^2,\epsilon) {\cal M}_0 + \overline{\cal M}_1.\label{m1}
\eeq
Hence to first order in $\alpha$ ($\alpha=e^2/4\pi$)
\beq
\delta = 2 f(Q^2,\epsilon) + 2 {\Re\{{\cal M}_0^\dagger
\overline{\cal M}_1\}\over |{\cal M}_0|^2}.\label{delta}
\eeq

The factorizable terms dominate, and include the electron vertex
correction, vacuum polarization, and the infrared (IR) divergent parts
of the proton vertex and two-photon exchange corrections. These terms
are all essentially independent of hadronic structure. The hadronic
model-dependent terms from the finite proton vertex and two-photon
exchange corrections are expressed in $\overline{\cal M}_1$. These
terms are small, and are generally ignored~\cite{MT69}. The finite
proton vertex correction was analyzed recently by Maximon and
Tjon~\cite{MT00}, who found $\delta<0.5\%$ for $Q^2<6$~GeV$^2$. It does
not show a significant $\epsilon$-dependence, and so we drop it here.

The factorizable terms can be classified further. Each of the functions
$f(Q^2,\epsilon)$ for the electron vertex, vacuum polarization, and
proton vertex terms depend only on $Q^2$, and therefore have no
relevance for the LT separation aside from an overall normalization
factor. Hence of the factorizable terms, only the IR divergent
two-photon exchange contributes to the $\epsilon$-dependence of the
virtual photon corrections.

For the inelastic bremsstrahlung cross section, the amplitude for real
photon emission can also be written in the form of Eq.\ (\ref{m1}). In
the soft photon approximation the amplitude is completely factorizable.
A significant $\epsilon$-dependence arises due to the frame-dependence
of the angular distribution of the emitted photon. These corrections,
together with external bremsstrahlung, contain the main
$\epsilon$-dependence of the radiative corrections, and are accounted
for in the experimental analyses~\cite{Wal94}.

\begin{figure}
\includegraphics[width=8cm]{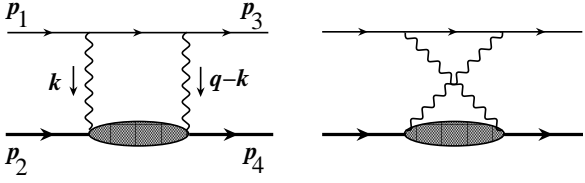}
\caption{Two-photon exchange box and crossed box diagrams.}
\label{fig1}
\end{figure}

In principle the two-photon exchange contribution to ${\cal M}_1$,
denoted ${\cal M}^{\gamma\gamma}$, includes all possible hadronic
intermediate states (Fig.~\ref{fig1}). Here we consider only the
elastic contribution to the full response function, and assume that the
proton propagates as a Dirac particle. We also assume that the
off-shell current operator is given by (\ref{current}), and use
phenomenological form factors at the $\gamma p$ vertices. Clearly this
creates a tautology, as the radiative corrections are also used to
determine the experimental form factors. However, because $\delta$ is a
ratio, the model-dependence cancels somewhat, provided we use the same
phenomenological form factors for both ${\cal M}_0$ and ${\cal
M}^{\gamma\gamma}$ in Eq.\ (\ref{delta}).

The sum of the two-photon exchange box and crossed box diagrams has
the form
\bea
{\cal M}^{\gamma\gamma} &=& e^4 \int {d^4 k\over (2\pi)^4}
\left[{N_a(k) \over D_a(k)} + {N_b(k) \over D_b(k)}\right],\label{mbox}
\eea
where the numerators are the matrix elements
\bea
N_a(k) &=& 
 \ubar(p_3) \gamma_\mu (\slash{p}_1 - \slash{k}) \gamma_\nu
 u(p_1)\nonumber\\
 &\times& \ubar(p_4) \Gamma^\mu(q-k) (\slash{p}_2 + \slash{k} + M)
 \Gamma^\nu(k) u(p_2),\label{nbox}\\
N_b(k) &=& 
 \ubar(p_3) \gamma_\nu (\slash{p}_3 + \slash{k}) \gamma_\mu
 u(p_1)\nonumber\\
 &\times& \ubar(p_4) \Gamma^\mu(q-k) (\slash{p}_2 + \slash{k} + M)
 \Gamma^\nu(k) u(p_2),\label{nxbox}
\eea
and the denominators are the products of the scalar propagators,
\bea
D_a(k) &=& [k^2-\lambda^2] [(k-q)^2-\lambda^2] \nonumber\\
&&\times [(p_1-k)^2-m^2] [(p_2+k)^2-M^2],\label{dbox}\\
D_b(k) &=& D_a(k) |_{p_1-k\rightarrow p_3+k}.\label{dxbox}
\eea
An infinitesimal photon mass $\lambda$ has been introduced in the
photon propagator to regulate the IR divergences, and the electron mass
$m$ is ignored in the numerator.

The implementation of Eq.\ (\ref{mbox}) is the main result of this
letter. However, we also want to compare with previous work, so a
partial analysis of the leading terms in (\ref{mbox}) is warranted.

To proceed, we can separate out the IR divergent parts from the finite
ones. There are two poles in the integrand of (\ref{mbox}) where the
photons are soft: one at $k=0$, and another at $k=q$. For the box
diagram, the matrix element can be written as the sum of a contribution
at the pole $k=0$ plus a remainder, $N_a(k) = N_a(0) +
\overline{N}_a(k)$. Explicitly, we have
\beq
N_a(0) = 4 \pepa q^2 i {\cal M}_0/e^2.
\eeq
The matrix element at the pole $k=q$ is the same, so $N_a(q)=N_a(0)$
(this also follows from symmetry arguments). This suggests that the
dominant contribution to the box amplitude can be approximated as
\beq
{\cal M}_a^{\gamma\gamma} \approx e^4 N_a(0) \int {d^4 k\over (2\pi)^4}
{1\over D_a(k)}\equiv {\cal M}_a^{\rm IR}.\label{mboxapp}
\eeq
There are two assumptions implicit in this approximation. The first is
that the integral involving $\overline{N}_a(k)$ is small, and contains no
ultraviolet (UV) divergences from the $F_2$ part of the current
operator (\ref{current}). Without hadronic form factors, Eq.\
(\ref{current}) does in fact lead to UV divergences. We demonstrate
below how to get around this difficulty by rewriting $F_1$ and $F_2$ in
terms of the Sachs form factors $G_E$ and $G_M$. The second assumption
is that the hadronic form factors have no significant effect on the
loop integral, and can be factored out. In essence, this assumes that
the hadronic current operators occurring in Eq.\ (\ref{mbox}) can be
replaced by $\Gamma^\mu(0)=\gamma^\mu$ for the vertex involving the
soft photon, and by $\Gamma^\mu(q)$ for the other vertex.

With these caveats in mind, the IR divergent box amplitude from the
pole terms can now be written as~\cite{MT00}
\bea
{\cal M}_a^{\rm IR} &=& {\alpha\over \pi} \pepa q^2 {\cal M}_0 
{i\over \pi^2} \int d^4 k\ {1\over D_a(k)}\nonumber\\
&=& - {\alpha\over\pi} \ln{\left(2\pepa\over m M\right)}
\ln{\left(Q^2\over \lambda^2\right)} {\cal M}_0.\label{mboxir}
\eea
The four-point function arising from the loop integral has been
evaluated analytically in the limit $\lambda^2\ll Q^2$ following
't~Hooft and Veltman~\cite{HV79}.

A similar analysis of the crossed box amplitude shows that
\beq
N_b(0) = 4 \pepb q^2 i {\cal M}_0/e^2,
\eeq
and hence
\beq
{\cal M}_b^{\rm IR} =
{\alpha\over\pi} \ln{\left(2\pepb\over m M\right)}
\ln{\left(Q^2\over \lambda^2\right)} {\cal M}_0.\label{mxboxir}
\eeq
In the lab frame ($\pepa=E_1 M$ and $\pepb=E_3 M$), the total IR
divergent two-photon exchange contribution to the cross section is
readily seen to be
\beq
\delta_{\rm IR} = -2{\alpha\over \pi} \ln{\left(E_1\over E_3\right)}
\ln{\left(Q^2\over\lambda^2\right)}, \label{deltair}
\eeq
a result given by Maximon and Tjon~\cite{MT00}. The logarithmic terms
in $m$ cancel in the sum, while the logarithmic IR singularity in
$\lambda$ is exactly cancelled by a corresponding term in the
bremsstrahlung cross section involving the interference between real
photon emission from the electron and from the proton.

By contrast, in the standard treatment of Mo and Tsai (MT)~\cite{MT69}
the loop integral in (\ref{mboxir}) is approximated by setting the
photon propagator not at a pole equal to $1/q^2$. This results in a
3-point function $K(-p_1,p_2)$ which, unfortunately, has no simple
analytic form in the limit $\lambda^2\ll Q^2$. After a further
approximation $K(-p_1,p_2)\approx K(p_1,p_2)$, the total IR divergent
result is given as~\cite{MT69}
\beq
\delta_{\rm IR}({\rm MT}) = -2{\alpha\over \pi} \left( K(p_1,p_2) -
K(p_3,p_2)\right), \label{deltairMT}
\eeq
where $K(p_i,p_j) = p_i\cdot p_j\,\int_0^1 dy\,\ln{(p_y^2/
\lambda^2)}/p_y^2$ and $p_y=p_i y + p_j (1-y)$.

\begin{figure}
\includegraphics[height=8cm,angle=90]{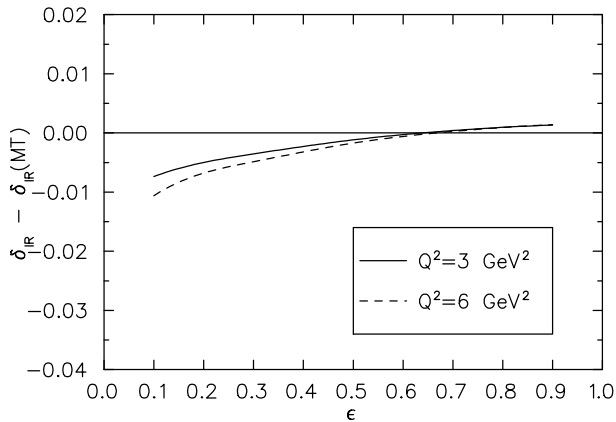}
\caption{Difference between the model-independent IR divergent contributions
of Eq.\ (\ref{deltair}) and of the commonly used expression
(\ref{deltairMT}).}
\label{fig2}
\end{figure}

Because $\delta_{\rm IR}({\rm MT})$ is the result generally used in
existing experimental analyses~\cite{Arr03,Wal94}, it is useful to
compare the $\epsilon$-dependence with that of $\delta_{\rm IR}$. The
difference $\delta_{\rm IR} - \delta_{\rm IR}({\rm MT})$ is independent
of $\lambda$, and is shown in Fig.~\ref{fig2} as a function of
$\epsilon$ for $Q^2=3$~GeV$^2$ and $Q^2=6$~GeV$^2$. The different
treatments of the IR divergent terms already have significance for the
LT separation, resulting in roughly a 1\% change in the cross section
over the range of $\epsilon$. This effect alone gives a reduction of
order 3\% and 7\% in the ratio $R$ for $Q^2=3$~GeV$^2$ and
$Q^2=6$~GeV$^2$, respectively.

We return now to the implementation of the full expression of Eq.\
(\ref{mbox}). The full expression includes both finite and IR divergent
terms (there is no need to treat them separately), and form factors at
the $\gamma p$ vertices. To avoid sensitivity to the UV divergences in
the loop integrals arising from the $F_2$ part of the current operator
(\ref{current}), we rewrite $F_1$ and $F_2$ in terms of the Sachs form
factors
\bea
F_1(q^2) &=& {G_E(q^2) + \tau G_M(q^2)\over 1+\tau},\label{f1tosachs}\\
F_2(q^2) &=& {G_M(q^2)-G_E(q^2)\over 1+\tau}.\label{f2tosachs}
\eea
$G_E$ and $G_M$ are taken to have the common form factor dependence
$G_E(q^2)=G_M(q^2)/\mu_p\equiv G(q^2)$, with $G(q^2)$ a simple monopole
$G(q^2)=-\Lambda^2/(q^2-\Lambda^2)$. We leave a fuller exploration of
the hadronic model-dependence to a future paper. Effectively the $F_2$
part of the current then behaves like a dipole, and the loop integrals
are UV finite for any choice of cutoff mass $\Lambda$. We have taken
$\Lambda=0.84$~GeV, consistent with the size of the nucleon, for which
the results show a plateau of stability. The sensitivity to $\Lambda$
is mild because the form factor dependence enters as a ratio in
$\delta$.

The loop integrals in Eq.\ (\ref{mbox}) can be evaluated analytically
in terms of four-point Passarino-Veltman functions~\cite{PV79}, and
trace techniques used to implement the sum over Dirac spinors implicit
in Eq.\ (\ref{delta}). This is a formidable task that is facilitated by
the use of established algebraic manipulation routines. We used two
independent packages (FeynCalc~\cite{feyncalc} and
FormCalc~\cite{formcalc}), which gave identical numerical results. The
Passarino-Veltman functions were evaluated numerically using the FF
program~\cite{ff}.

\begin{figure}
\includegraphics[height=8cm,angle=90]{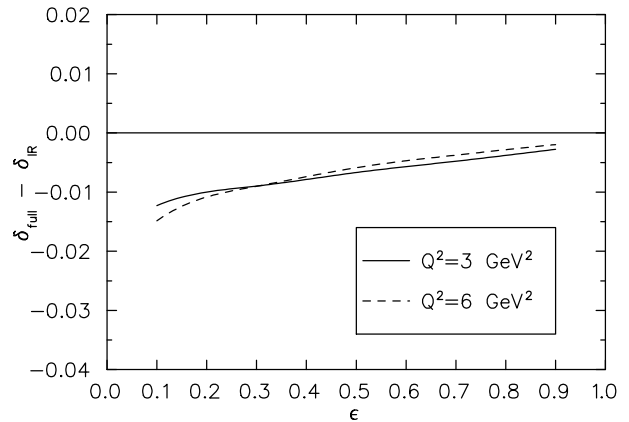}
\caption{Difference between the full two-photon exchange correction and
the model-independent IR divergent result of Eq.\ (\ref{deltair}).}
\label{fig3}
\end{figure}

The model-independent IR divergent result of Eq.\ (\ref{deltair}) is an
appropriate benchmark with which to compare the full result
$\delta_{\rm full}$. Because the IR behavior is the same, the
difference $\delta_{\rm full}-\delta_{\rm IR}$ is finite (i.e.\
independent of $\lambda$). The results are shown in Fig.~\ref{fig3}. A
significant $\epsilon$-dependence is observed, which increases slightly
with $Q^2$. The additional correction is largest at backward angles
($\epsilon\rightarrow 0$), and essentially vanishes at forward angles
($\epsilon\rightarrow 1$).

To consider the effect on the ratio $R$ determined in the LT
separation, we make a simplified analysis that assumes the modified
cross section is still approximately linear in $\epsilon$. The results
shown in Figs.~\ref{fig2} and \ref{fig3} are combined, giving
$\Delta=\delta_{\rm full}-\delta_{\rm IR}({\rm MT})$. For each value of
$Q^2$ in the range 1-6~GeV$^2$ we fit the correction $(1+\Delta)$ to a
linear function of $\epsilon$ of the form $a (1+b\epsilon)$. The
parameter $b$ so determined behaves roughly like $b\approx 0.014
\ln{(Q^2/0.65)}$, with $Q^2$ in GeV$^2$. For the LT separation, the
corrected Eq.\ (\ref{sigma0}) becomes
\beq
d\sigma = (a A) \tau G_M^2(Q^2) \left(1 + (B \widetilde{R}^2 +
b)\epsilon\right),
\eeq
where $B=1/(\mu_p^2\tau)$, and $\widetilde{R}$ is the corrected ratio $R$.
Since $a\approx 1$, we have essentially $\widetilde{R}^2 = R^2 - b/B$.

\begin{figure}
\includegraphics[width=8cm]{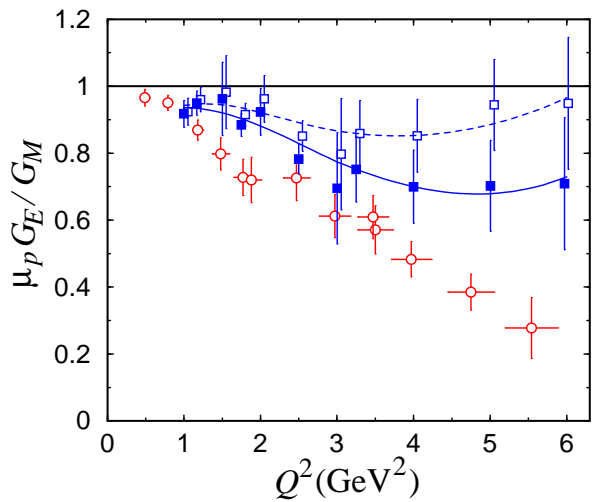}
\caption{The ratio of form factors measured using LT separation (hollow
squares), together with the global fit (dashed line). The unshifted LT
data represent a binned average of all LT separated data points with
normalization factors determined by the global fit in
Ref.~\cite{Arr03}. Filled squares show the shift in the LT results due
to the two-photon exchange corrections (offset for clarity), and the
solid line shows the effect on the global fit. Error bars have been
left unchanged. The polarization transfer data~\cite{Jon00} are shown
as hollow circles.}
\label{fig4}
\end{figure}

The shift in $R$ is shown in Fig.~\ref{fig4}, together with the
polarization transfer data. The effect of the additional terms is
significant. Although some dependence on nucleon structure is expected,
these calculations show that the two-photon corrections have the proper
sign and magnitude to resolve a large part of the discrepancy between
the two experimental techniques. Clearly there is room for additional
contributions from inelastic nucleon excitation (e.g. the $\Delta^+$).
These have been examined previously in Refs.~\cite{DR57} in various
approximations. Greenhut~\cite{DR57} used a fit to proton Compton
scattering to calculate the resonant contribution to two-photon
exchange, and found some degree of cancellation with the nonresonant
terms at high energies. Further study of the inelastic region is
required, including also the imaginary part of the response
function~\cite{Afa02}.

Direct experimental evidence for the contribution of the real part of two-photon
exchange can be obtained by comparing $e^+p$ and $e^-p$ cross sections.
(${\cal M}_0$ changes sign under $e^-\rightarrow e^+$, whereas ${\cal
M}^{\gamma\gamma}$ does not.) Hence we expect to see an enhancement of
the ratio $\sigma(e^+p)/\sigma(e^-p)$ due to two-photon exchange (after
the appropriate IR divergences are cancelled due to bremsstrahlung).
There are experimental constraints from data taken at SLAC~\cite{Mar68}
for $E_1=4$~GeV and $E_1=10$~GeV, which are consistent with our
results. However, the SLAC data are from forward scattering angles,
with $\epsilon>0.72$, where we find the two-photon exchange
contribution is $\alt 1\%$. A more definitive test of the two-photon
exchange mechanism could be obtained at backward angles, where an
enhancement of order a few percent is predicted.

\acknowledgments
We thank A.~Afanasev, J.~Arrington, S.~Brodsky, K.~de~Jager, and
R.~Segel for helpful discussions. PGB also thanks the theory group and
Hall C at Jefferson Lab for support during a sabbatical leave, where
this project was undertaken. This work was supported in part by NSERC
(Canada), DOE grant DE-FG02-93ER-40762, and DOE contract
DE-AC05-84ER-40150 under which SURA operates the Thomas Jefferson
National Accelerator Facility.

\end{document}